\documentstyle[aps,12pt]{revtex}

\begin{document}
\title{Magnetostriction Measurements with Atomic Force Microscopy: A Novel Approach}
\author{A. C. Papageorgopoulos, H. Wang, C. Guerrero and N. Garcia*}
\address{Laboratorio de Fisica de Sistemas Peque\~{n}os y Nanotechnologia CSIC,\\
Serrano\\
144 E-28006 Madrid, Spain}
\maketitle

\begin{abstract}
In this study we present a new method of measuring magnetostriction with an
atomic force microscope adapted for the application magnetic fields. The
experiment allows us to visualise, in an elegant and educational way how the
lateral magnetoelastic shape changes take place on the sample surface when a
magnetic field is applied. We have, furthermore, used this technique to
observe magnetically induced strains as small as $5\cdot 10^{-8}$, and have
measured Ni, permalloy and commercial Cu wires and films, as well as pure Cu
and Pt wires, where results are in agreement with other methods of
measurement. The applications are, moreover, relevant to studies of
ballistic magnetoresistance, where we can draw conclusions involving the
effect of the magnetically induced strains on magnetoresistance measured at
the same time as magnestostriction.



* corresponding author

email: nicolas.garcia@fsp.csic.es

Tel: \ 34-91-561 88 06

Fax: 34-91-563 15 60
\end{abstract}

\bigskip \newpage\ \ \ \ \ Magnetostriction is the phenomenon whereby the
shape of a ferromagnetic specimen changes during the process of
magnetization.\cite{1} The deformation $\Delta $l/l resulting from this
change is usually in the 10$^{-5}$ to 10$^{-6}$ range. Magnetostriction can
be positive or negative depending if the material expands or contracts along
the measured direction.\cite{2}  Many ways exist for measuring this
field-induced deformation, including the strain gauge and capacitance
methods.\cite{1,2,3,4,5} The development of local surface probes (scanning
tunnelling microcopy (STM), scanning force microscopy (SFM), magnetic field
microscopy (MFM) and others) \cite{6,7}, however, has introduced new ways to
visualize the structure and topology of the surfaces with high spatial
resolution. These surface characterization techniques can easily visualize
variations of the surface structure as well as local surface behaviour such
as those resulting from induced strains on the measured specimen. Takata $et$
$al$, in particular, have reported strain imaging of magnetic recording
material.\cite{8} By their definition, however, strain imaging involves
detecting strains generated by any modulation including an alternating
magnetic field. In the case of Ref. 8, the strain is driven in the
z-direction by the alternating B-field. Up to now, and according to our
knowledge, there have never been applications of local probe techniques to
observe and measure $lateral$ changes of the shape of a specimen under an
applied magnetic field $without$ $modulation$. Given the vast applications
of magnetic materials, the need for magnetic characterization methods of
greater resolution is paramount. We, therefore, believe that carefully
set-up experiments of this nature should provide very valuable information.

On another note, ever since preliminary experiments showing large ballistic
magnetoresistance (BMR) effects in atomic nanocontacts have been reproduced
in other cases,\cite{9,10,11} there have been observations of BMR values of
700\%,\cite{12} 3000\% and practically infinity.\cite{13} The latter are
observed when Ni contacts of nanometer size are electrodeposited in the gap
region of Ni wires. In addition, BMR values of over 50000\% have been
observed when using permalloy Fe$_{21}$Ni$_{79}$ wires, and values not as
large (a few hundred percent) \cite{14} were measured in commercial Cu
wires. From the above, a very legitimate question can be posed: $What$ $is$ $%
the$ $effect$ $of$ $the$ $magnetostriction$ $deformation$ $of$ $the$ $%
nanocontacts$ $on$ $the$ $BMR?$ $In$ $other$ $words,$ $is$ $the$ $BMR$ $due$ 
$to$ $the$ $magnetoelastic$ $deformations?$

In this paper we present a new method of measuring magnetostriction with the
use of atomic force microscopy (AFM). With the AFM we are able to visualize
in an illustrative and educational manner the strains of a ferromagnetic
sample's micro and nanostructure under an applied magnetic field. In our
technique, the field is unidirectional for each application with the
intended value applied near-instantly and removed in the same manner after a
period of influence lasting several seconds (and scan lines). There is no
field modulation, and the lateral shift is measured directly from the scan.
Lateral strains as small as $5\cdot 10^{-8}$ have been thus measured,
although this resolution could be improved under more stringent scanning
conditions. Sample size, however, is not an issue as long as a topographic
image can be attained. Thus, the technique is easily applicable to samples
whose length is of the order 100 $\mu $m as well as larger specimens. The
wires measured in this study included Ni, permalloy, commercial and pure Cu
as well as Pt. In addition, applications of this method are performed to
answer the question posed above by measuring BMR of nanocontacts formed via
electrodeposition on wires. From our results: we do not see a direct
relation between magnetostrictive strains and large BMR.

The experiment was conducted with a Dimension 3100 Scanning probe Microscope
(SPM), with a Nanoscope IV Controller. Figure 1(a) shows a diagram of the
experimental set-up during measurements. The U-shaped electromagnet
(constructed in the laboratory for this purpose) was comprised of a ZnMn
ferrite powder core, wrapped with Cu wire on one end. The magnet was,
furthermore, supported so the poles could be accurately placed on either
side of the sample (without coming into contact with the latter), with the
desired stability during scanning. This configuration provided a maximum
field of about 250 Oe (measured with a DTM-133 Teslameter at the sample
plane) for these measurements. By controlling the current of the 75 W, DC
power supply, the magnetic field strength could be varied to within 
\mbox{$<$}%
2 Oe.

The metal wires used were Ni, permalloy Fe$_{21}$Ni$_{79}$, Cu (both
commercial wire and that of 99.999\% purity) and Pt. With the exception of
Ni, with a diameter d = 0.25 mm, all the metal wires were of d = 0.5 mm.
Fig.1(b) shows, in diagrammatic form, how the wires were attached to their
base (made of circuit-board material) for both the cases of free wires and
wires in a ``T''-configured contact formation (upper and lower diagrams of
fig. 1(b) respectively). Cyanoacrylic commercial glue was the preferred
sample clamping method, used successfully in all cases. The glue was applied
to a length l$^{2}$ at one end of the wire, leaving a predetermined length l$%
^{1}$ free. The distances from the glue boundary to the AFM tip for each
wire were measured using an OLYMPUS BX50 optical microscope in combination
with the built-in optical microscope camera of the Dimension 3100 SPM.

Figure 1(c), shows a side-view close-up of the scanning process. Contact
mode AFM was always used, with short (100 $\mu $m) cantilevers with a force
constant of 0.58 N/m being preferred. Fig. 1(c) schematically depicts the
case of a free wire measured under a field direction parallel to its length.
The scan direction was always along the length of the wire, whereupon
changes in the length were easily detected as shifts in the x-direction of
the scan image.

$Ni$ $free$ $wires$ $and$ $``T"-configured$ $contacts$: Our first
measurements were conducted on Ni, a ferromagnetic material with well-known
magnetoelastic properties. Wires were measured in two configurations: a)
single free wires and b) two wires forming a ``T'' configuration (fig. 1(b),
lower portion). Results are depicted in fig. 2. Specifically, the upper
portion of fig. 2 shows two typical scans of the Ni wire surface (left and
center) using contact mode AFM, and an optical microscope close-up image of
the measured ``T''-configured contact (right). The two scans are both 300 nm
in range with the positive x-direction toward the right of each scan. In the
case of the far left scan the magnetic field (H= 80 Oe) was consecutively
applied and removed five times. The scan direction (as indicated by the
vertical arrow) was from the bottom to the top of the image, and the first
application of the field (toward the bottom of the scan), including the
direction of strain, is depicted by the black arrow pointing left
(indicating contraction). In contrast, the bottom portion of the figure
shows the arrow pointing right, which is indicative of the field being
turned off (re-expansion to original length). As can be easily seen in the
far left-hand scan of fig. 2, the shift is nearly instantaneous whether the
field is applied or removed, i.e., there are no distortions of the scan
image during magnetic field transitions. The distance of the shift was
easily measurable using available SPM software, and the repeatability of the
effect is easily demonstrated by the identical multiple shifts in the scan.
Another important result is that when the field is removed the wire returns
to precisely its original position as indicated by the topographic
continuity observable in the far left image between field applications. The
contrast between the far left-hand scan and that at the upper middle of Fig.
2, where no field is applied is obvious, although the scans do not show
identical topographic features (scan windows are within 1 m of each other on
the surface). Even with images with a great degree of noise, the effects of
magnetoelastic strain are easily distinguishable from the former at least
for strains down to $5\cdot 10^{-8}$.

The bottom portion of fig.2 depicts the results of various measurements of
different configurations of Ni wires under the influence of both parallel
and transversal fields, where changes in field intensity reach 250 Oe. The
upper portion of the graph (fig. 2(a)) depicts the changes in length of a 6
mm free wire during the indicated field application when field direction is
transversal to the length of the wire, while the lower portion (fig. 2(b))
shows the magnetostiction of a 6mm free wire when the same field values are
applied in the direction parallel to the wire's length. The extra data
points on the right-hand side of both plots represent singular values of
wire deformation measured under different sample mounting and field
direction conditions. The plot (represented with inverted solid triangles)
in fig.2 (b), represents the deformation of the ``T''-configured Ni contact
vs. increasing field strength. First, the plots of the deformation of the
free wire vs. field strength for magnetic field directions both parallel and
transversal to the wire lengths have been presented elsewhere, \cite{15} and
are in agreement with previous findings for the magnetoelastic deformation
of Ni wires under a field direction parallel to their length.\cite{1,2,3}
Second, the question whether the glue itself provides adequate clamping to
prevent movement of the wire as a whole when under a magnetic field, and
thus eliminates the strain effects caused by the total magnetization of the
specimen was addressed by first measuring the strain on a wire where the
glued portion was five times the length of the free (1 mm) wire. We found
the deformation to be twice that of a wire with only 25\% of its total
length glued (such as the wires used in the plots of Fig. 2 (a) and (b)). We
then eliminated the strain contribution of the glued portion by measuring
the magnetostriction at two points on the free wire, and applying the
formula $\ \left( \Delta l_{2}-\Delta l_{1}\right) /\left(
l_{2}-l_{1}\right) $, where $l_{2}$ and $l_{1}$ are the free wire lengths
corresponding to the two positions. Results deviated only 5\% from the case
of the wires glued along 25\% of their total length, and could be improved
with a smaller percentage of the total wire length glued. For a more
detailed explanation see Ref. 15.

Our measurements continued with the magnetostriction of a 6 mm free Ni wire
(only 2mm glued) in a ``T''-configured contact formation, with a contact
area measured at 30 m. The upper right-hand portion of figure 2 shows a 600
m Dimension 3100 optical microscope picture of the ``T'' contact area. The
scans were performed 200 $\mu $m from the contact itself. With these
measurements along with those on permalloy wire we attempt to shed some
light on the relationship between magnetoelastic deformation in
``T''-configured contacts and BMR. For this reason, the resistance was
monitored over the contact throughout all the magnetic field applications
involving ``T''-configured contacts. The value of this resistance was
measured at 1.5 $\Omega $%
\c{}%
, which represents the effective conductive portion of the contact. From
previous studies we have found this particular value to correspond to
conducting contact of 30 nm,\cite{14} in contrast with the total contact
geometry (30 $\mu $m diameter area). Most of the total contact area is
actually comprised of non-conductive oxides.

In fig. 2(b) the upper curve represents the deformation values of the
``T''-contact vs. increasing magnetic field, up to 250 Oe in a direction
parallel to its length. The observed deformation at 250 Oe is about half the
value of the free wire (-16 nm/mm) when the field is parallel, and about 2.5
nm/mm when it is transversal to the wire's length (represented by the hollow
upright triangle in fig. 2(a)). In the case of the parallel field direction
in particular, it would be expected that a 100 nm contraction (over the 6 mm
free length) would break the contact. In fact, the monitored resistance
remained stable during multiple consecutive field applications. In other
words, magnetoelastic strain does not alter the resistance across the
contact, and more importantly does not automatically imply large
magnetoresistance. It is most likely that both wires of the
``T''-configuration in the contact area deform together when the field is
applied. In other words, we must consider the contact connecting the two
wires as a single system extending to the second wire.

It is known that 10\% of the contact samples exhibit magnetoresistance in
the 100\% range and only 2\% of the samples show BMR values over 1000\% \cite
{12,13,14}. All ferromagnetic materials exhibit some degree of
magnetoelastic deformation, including those exhibiting BMR. Although it is
evident that magnetostriction does not necessarily imply BMR, we can draw no
conclusions involving the reverse from the available data on Ni. But there
is no evidence that the contacts are modified as indicated from our
resitance and magnetostriction data. In other words, based on the above
results, we cannot say one way or the other if BMR induces magnetoelastic
strain.

\bigskip

$Permalloy$ $Fe_{21}Ni_{79}$ $wires$: Similar measurements were conducted on
permalloy Fe$_{21}$Ni$_{79}$, although only for the cases of the free wire
and ''T''-configured contact both under the influence of an applied field
parallel to the wire lengths. Figure 3 shows a similar configuration as that
of fig. 2, this time of the permalloy ``T''-configured contact. The top
portion of the figure shows two scans of 450 nm of the wire surface, about
200 $\mu $m from the contact tip. The left-hand scan involves magnetic field
applications, and the subsequent respective topographic shifts in the image,
while the right-hand scan represents the surface without any magnetic field
applied. Below these images is a graph showing the changes in length of a 10
mm permalloy Fe$_{21}$Ni$_{79}$ wire when increasing field values are
applied in the parallel direction. Our measured contraction of permalloy
wire with increasing field strength is approximately equal to that indicated
in the literature,\cite{2} where $\Delta $l/l approaches -2 nm per mm length
of wire at higher field values (H $\geq $80 Oe). To examine the case for BMR
applications, an additional data point (hollow circle) was added. This
represents the measured contraction of the permalloy wire under an 80 Oe
parallel field, when forming a contact (in a ''T''-configuration as with
Ni). The permalloy contact also remained intact with a stable resistance
during deformation. The obtained value for the permalloy contact was,
furthermore, practically identical to that of the free permalloy wire. In
the case of permalloy Fe$_{21}$Ni$_{79}$, however, the movement is too small
to meet resistance from the glue or transversal wire. It is most likely for
this reason that we do not observe any decrease in the magnetostriction
values of permalloy when comparing the field induced strain of the
''T''-configured contact with that of the free wire.

$Copper$ $and$ $Platinum$ $wires$: We conducted measurements on three
paramagnetic wires, each 10 mm long with applied fields in directions
parallel to their respective lengths, and have provided some interesting
results. Commercial Cu was measured first, initially as a reference sample
to the Ni and permalloy wires. Despite the fact that Cu is a paramagnetic
material, there was magnetostrictive strain that increased with increasing
magnetic field strength, although in a more linear manner than Ni and
permalloy. The changes in length measured in the commercial Cu sample,
moreover, were of the same order as permalloy wires under a parallel field.
Figure 4 (top left) shows a case where a field is applied to commercial Cu
with the evident shift (in the same manner as the previous ferromagnetic
samples under parallel fields). The top right part of the figure depicts the
plot of the contraction of the commercial Cu wire with increasing field
strength, as described above. Our commercial sample was analysed by EDAX
where the presence of 3\% Co, Ni and Fe impurities was detected. It is
evident from this analysis that even minute amounts of impurities cause
appreciable changes in the values of magnetoelastic strain in a paramagnetic
material such as Cu. To compare we have attempted to measure
magnetostriction in pure (99.999\%) Cu, as well as Pt wires (not shown
here), and the scans do not exhibit any observable change under the same
applied field (250 Oe). I.e. there is a complete absence of the
characteristic topographic displacement in the x-scan direction present in
measurements of ferromagnetic specimens. It should be noted that pure Cu
actually exhibits a magnetostriction deformation of 10$^{-9}$, which for a
10 mm of Cu wire implies a displacement of 0.01nm (0.1 \AA ) \cite{16}. At
the bottom right-hand of the figure is a scanning electron micrograph (SEM)
of a Cu film with a 10 $\mu $m gap, bridged by electrodeposited permalloy,
where we measured for BMR and magnetostriction. While close to 100\% BMR was
measured here, (repeatable through several hundred trials), there was no
observable magnetostriction under applied fields up to 250 Oe. The lack of
displacement described for the pure Cu and Pt samples is clearly shown in
the 200 nm scan of the aforementioned contact (fig.4, bottom left), where
repetitive field applications were performed while scanning under the AFM.
It is no surprise, regarding the above information, that a pure Cu film
shows no magnetostriction. The permalloy deposite should, however, also be
taken into consideration. As we have described, permalloy Fe$_{21}$Ni$_{79}$
exhibits a magnetostrictive strain around 10$^{-6}$, which implies a strain
of 0.01 nm (0.1 \AA ) for a 10 $\mu $m length of the material. We have
attempted to see this displacement with an ex-situ STM, but resolution was
in the Angstrom range, and no topographic shifting was detected. It is
noteworthy that even a displacement below what is the minimum detectable
causes a disturbance in the scan that manifests as a horizontal line through
the image at the point the field is applied. Whether, however, this 0.1\AA\ %
displacement of the contact can influence the BMR response can be understood
as follows: Our 1-10 Ohm contacts correspond to sections of contacts of 10$%
^{3}$ to 10$^{4}$ atoms- taking one atom to be one conductance unit for a
good conductor (for a bad one the section is larger still). In order to
obtain the observed changes of 200\% such as those described in the case
below, we would need to change the section by a factor of 3. This, according
to all simulations regarding the pulling of nanowires, is not possible for a
shift of 0.1 \AA .

We thought it relevant to conclude this report with a brief description of
our most recent magnetostriction data (not shown here), involving the
measurement of a contact formed by the electrodeposition of permalloy to
bridge the 30 $\mu $m gap between two pure Cu wires aligned tip-to-tip \cite
{17}. The applied field strength in this case was H = 850 Oe, and no shift
was observed in the scan upon field application, meaning that if there was a
strain present it was 
\mbox{$<$}%
$5\cdot 10^{-8}$. As described in the former paragraph, such a strain cannot
correspond to the measured $BMR$ $of$ $200\%$ in this system. From the very
fact that we did observe BMR in at least these two latter cases, we may
conclude that not only does magnetostriction not imply BMR, but that also
the presence of BMR does not mean observable magnetostriction is present.
The two effects are thus not related to each other.

In this report a new method of measuring the magnetostriction of metallic
wires has been presented, which utilizes atomic force microscopy (AFM), and
the (near-instantaneous) application of a magnetic field at the sample
plane. The strains are observed laterally, relative to the sample plane, in
the direction of the scan in progress. This technique eliminates the
contribution of strains due to magnetization of the total specimen, and the
lack of modulation guarantees the absence of electromagnetic effects due to
eddy currents caused by the effects of the oscillating magnetic field on the
sample. We have measured strains as small as $5\cdot 10^{-8}$ in sample
areas in scan ranges as small as 200 nm. The experiment shows in a dramatic
visual manner how the magnetoelastic strains of the sample take place when
the magnetic field is applied. For wires exhibiting magnetostriction, the
shift is instantaneous and is clearly depicted in the scanned images
presented in this study. Applications of this method have been made to wires
of Ni, permalloy, commercial Cu wire, pure Cu, and pure Pt, as well as
``T''-configured contacts of Ni and permalloy Fe$_{21}$Ni$_{79}$. These
applications are relevant in examining the effect of magnetoelastic strains
on magnetoresistance, measured at the same time as the magnestostriction.
From the data presented, there is no evidence that magnetostriction
automatically implies magnetoresitance. While large magnetostriction is seen
for all Ni samples, magnetoresitance is only seen for the 10\% of the
samples and only 2\% for BMR larger than 1000\%. This happens for Ni and
permalloy samples even if the deformations of the latter are 30 times
smaller than those of the former. We speculate, as discussed above, that the
contact moves rigidly with the magnetoelastic motion of the wires (as if the
contact and second wire are a continuation of the first). Further
measurements on permalloy contacts deposited on pure Cu films and wires,
which have exhibited up to 200\% BMR have shown no observable
magnetostriction. We, therefore, conclude that BMR and magnetostriction are
not causally linked, i.e. one effect does not imply the other.

\bigskip

Acknowledgements

We would like to extend our thanks to Prof. A. del Moral for his most
valuable advice and discussions, and also to N. D. Nikolic and H. Cheng for
their invaluable assistence in performing this experiment. This work has
been supported by the DGICyT.

\bigskip

\bigskip

\bigskip

$\bigskip $

$References:$

\newpage

$\bigskip $

$Figure$ $Captions$:

Figure 1: Schematic depicting the experimental set-up with (a) a top-view
diagram of the Dimension 3100 SPM with the sample holding system placed
between the poles of the electromagnet inserted under the scanner (shown
transparent). (b) Top-views of free wire and contact configurations, and (c)
a side-view schematic of the placement of the AFM tip as it scans the wire.

Figure 2: top-left and center: Two scans of the Ni free wire with a magnetic
field, applied multiple times throughout the scan, (left) and without any
field applied (right). In both cases the scan range is 300 nm, while the
scan direction is indicated by the red and black arrows in the corresponding
figures; (top right): Dimension 3100 optical microscope image of Ni
''T''-configured contact after parallel field applications. Scale is in m.
Contact area has been measured at 30 $\mu $m. (Bottom): Graphs of the
measured change in wire length $\Delta $l/l x10$^{6}$ vs. the strength H (in
Oe) of the applied field when (a) the field direction is transversal and
(b), when it is parallel to the wire's length. Upper plot of (b) refers to
Ni contact and the lower graph to the free wire. Additional data points mark
cases of magnetostriction for a contact under a transversal field (upright
hollow triangle), and the case of a 1mm free wire under a parallel field
(filled circle).

Figure 3: Top: two 450 nm contact AFM scans of permalloy Fe$_{21}$Ni$_{79}$
wire in ''T''- contact formation, with free length l = 10mm. Left scan shows
the application of a magnetic field, while right scan depicts the same
topography without any field applied. Bottom: Graph of $\Delta $l/l x10$^{6}$
vs. H with applied parallel field onto 10 mm free wire. Extra data point at
H = 80 Oe (enlarged hollow circle) represents permalloy ''T''-contact
deformation at this field strength.

Figure 4: Top row (left): 200 nm scan of a commercial Cu wire with a free
(unclamped) length of 10 mm with an applied filed of H = 250 Oe. Topographic
displacement to the left upon field application is obvious. Top row (right):
graph of $\Delta $l/l x10$^{7}$ vs. H (in Oe), with various field strengths
applied in the parallel direction. Bottom left: 200 nm scan of permalloy
electrodeposited over a 10 m gap in a pure Cu film. The scan was taken at a
distance of \symbol{126}200 $\mu $m from the center of the gap. Even with
multiple applications of a 250 Oe field, there was no magnetostriction
evident. Bottom right: SEM photo of the latter surface showing a close-up of
the contact area formed by electrodeposited permalloy. The 5 $\mu $m
scale-bar is placed vertically to the right of the contact. This specimen,
while showing no magnetoelastic response did exhibit a 100\% BMR (both were
measured at the same time).

\end{document}